\documentclass[12pt]{article}
\catcode`\@=11
\@addtoreset{equation}{section}

\global\arraycolsep=2pt
\usepackage{mathrsfs,amsbsy,amssymb,latexsym,amsfonts,amsmath,cite}
\usepackage{graphicx,color}

\newcommand\no{\nonumber}

\newcommand{\dd}{{\rm d}}
\newcommand{\omegati}{ \tilde{\omega}}

\allowdisplaybreaks

\begin{document}

\begin{flushright}
\parbox{4cm}
{KUNS-2715 \\ 
}
\end{flushright}

\vspace*{0.5cm}

\begin{center}
{\Large \bf Weyl transformation and regular solutions \vspace*{0.1cm}\\ 
in a deformed Jackiw-Teitelboim model
}
\vspace*{1.5cm}\\
{\large  Suguru Okumura\footnote{E-mail:~s.okumura@gauge.scphys.kyoto-u.ac.jp}
and Kentaroh Yoshida\footnote{E-mail:~kyoshida@gauge.scphys.kyoto-u.ac.jp}} 
\end{center}

\vspace*{1cm}

\begin{center}
{\it Department of Physics, Kyoto University, \\ 
Kitashirakawa Oiwake-cho, Kyoto 606-8502, Japan} 
\end{center}

\vspace{1cm}

\begin{abstract}

We revisit a deformed Jackiw-Teitelboim model with a hyperbolic dilaton potential, 
constructed in the preceding work [arXiv:1701.06340]. Several solutions are discussed  
in a series of the subsequent papers, but all of them are pathological because of a naked singularity 
intrinsic to the deformation. In this paper, by employing a Weyl transformation to 
the original deformed model, we consider a Liouville-type potential 
with a cosmological constant term. 
Then regular solutions can be constructed with coupling to a conformal matter   
by using $SL(2)$ transformations. 
For a black hole solution, the Bekenstein-Hawking entropy is computed from the area law. 
It can also be reproduced by evaluating the boundary stress tensor 
with an appropriate local counter-term 
(which is essentially provided by a Liouville-type potential). 

\end{abstract}

\setcounter{footnote}{0}
\setcounter{page}{0}
\thispagestyle{empty}

\newpage

\tableofcontents

\section{Introduction}

A recent interest in the study of String Theory is to establish a toy model of 
the AdS/CFT correspondence \cite{M,GKP,Witten}. In particular, it is significant 
to understand a quantum mechanical description of black hole with the holographic principle \cite{Hooft,Susskind}. 
Along this direction, 
Kitaev proposed a one-dimensional system composed of $N \gg 1$ fermions 
with a random, all-to-all quartic coupling \cite{Kitaev}. 
This model is a variant of the Sachdev-Ye (SY) model \cite{SY}, 
and so it is called the Sachdev-Ye-Kitaev (SYK) model. 
A remarkable point is that this model exhibits the maximal Lyapunov exponent 
in out-of-time-order four-point functions \cite{Kitaev,PR,MS}. 
Hence the SYK model may have a gravity dual described by an Einstein gravity 
\cite{SS1,SS2}\footnote{As another interesting direction, the conformal SYK model is proposed 
by Gross and Rosenhaus \cite{cSYK1,cSYK2}. Then, the bulk dual may not be a gravitational theory 
but scalar field theories on the rigid AdS$_2$\,.}. There are a lot of developments. 
All-point correlation functions have been computed in \cite{GR2}. The SYK model  
is extended to multi-flavor cases \cite{GR1}. Supersymmetric extensions are presented 
in \cite{FGMS}. A possible relation to 3D bulk dual is argued in \cite{Das}. 
The spectral density of the SYK model is analytically computed in \cite{Garcia}. Some models without 
the disordered coupling are proposed in \cite{Witten2,Nishinaka} and the related tensor models are discussed 
in \cite{Klebanov, Itoyama}. 

\medskip

A simple candidate of the gravity dual for the SYK model is a particular 1+1 dimensional dilaton 
gravity system originally introduced by Jackiw \cite{Jackiw} and Teitelboim \cite{Teitelboim} 
(called the Jackiw-Teitelboim (JT) model)\footnote{For a nice review on 2D dilaton-gravity, 
see \cite{V}.}. This model has been revisited by Almheiri and Polchinski \cite{AP} 
from the point of view of holography. Hence this model is sometimes called 
the AP model. A lot of efforts have been made to clarify the relation between the JT model 
and the SYK model, but it would be fair to say that the two models coincide in a low-energy region 
as the Schwarzian theory \cite{Jensen,EMV,MSY,SW}. 

\medskip

In the preceding works \cite{KOY,KOYmatter}, we have studied a deformation of the JT model 
by applying the Yang-Baxter deformation technique \cite{Klimcik,DMV,KMY}.
The deformed model has a hyperbolic dilaton potential. We have shown that solutions 
in the deformed model can be represented by a couple of Liouville's solutions 
and found several solutions such as the general vacuum solutions, 
shock waves and deformed black holes. Typically, the region near the boundary 
is deformed to a two-dimensional de Sitter (dS$_2$) and a new naked singularity appears. 
These are common features in geometries generated by Yang-Baxter deformations with 
classical $r$-matrices of Drinfeld-Jimbo type.
However, what these features indicate has not been revealed so far.

\medskip

In this paper, we study the deformed JT model by employing a proper frame 
proposed by Frolov and Zelnikov \cite{Frolov}. This frame is realized by performing 
a dilaton-dependent Weyl transformation and in this frame the deformed model is described 
as a Liouville gravity model with a cosmological constant. 
Then we find regular solutions with no singularities intrinsic to the deformations 
by using $SL(2)$ transformations. Among them, a black hole solution is included. 
We investigate the thermodynamics properties by examining the Bekenstein-Hawking entropy. 
This entropy can also be reproduced by evaluating the boundary stress tensor with 
a local counter-term. Notably, the counter-term in this proper frame 
is much simpler than the ones utilized in the preceding works \cite{KOY,KOYmatter}. 

\medskip 

This paper is organized as follows. Section 2 gives a brief review of the deformed JT model.
In section 3, we revisit the deformed JT model by employing a particular Weyl transformation. 
In this new frame, the deformed model is described as a Liouville gravity 
with a cosmological constant. 
In section 4, we consider a black hole solution and examine its thermodynamic properties. In particular, 
the Bekenstein-Hawking entropy is reproduced by evaluating the boundary stress tensor 
with an appropriate counter-term. Section 5 is devoted to conclusion and discussion.
In Appendix A, we show the difference between the black hole solution newly obtained 
in this paper and the one previously found in the preceding works \cite{KOY,KOYmatter}.

\section{A review of a deformed Jackiw-Teitelboim model}
 
In this section, let us give a short review of a deformed Jackiw-Teitelboim (JT) model 
with a hyperbolic dilaton potential presented in the preceding works  \cite{KOY,KOYmatter}. 

\subsection{The general vacuum solutions}

We work here in the Lorentzian signature and 
the (1+1)-dimensional spacetime is described by the coordinates 
$x^{\mu}=(x^0,x^1)=(t,x)$\,. The basic ingredients of this system are the 2D metric $g_{\mu\nu}$ 
and the dilaton $\Phi$\,. 

\medskip

The classical action for $g_{\mu\nu}$ and $\Phi$ is given by \cite{KOY,KOYmatter}\footnote{
The sinh-type potential is related to a $q$-deformed $\mathfrak{sl}(2)$ algebra 
via (4.1) in the work \cite{II}.} 
\begin{eqnarray}
S_{\Phi } &=& \frac{1}{16\pi G}\int\!\dd^2 x\,\sqrt{-g}\left[\Phi^2 R + \frac{1}{\eta L^2}\sinh\left( 
2 \eta\, \Phi^2 \right)\right] +\frac{1}{8 \pi G}\int\!\dd t\, 
\sqrt{-\gamma_{tt}}\,\Phi^2 K\,, 
\label{sinh action}
\end{eqnarray}
where $G$ is a two-dimensional Newton constant, $R$ and $g$ are Ricci scalar 
and determinant of $g_{\mu\nu}$\,, and $L$ is an AdS radius.
The last term is the Gibbons-Hawking term that consists of extrinsic metric $\gamma_{tt}$ 
and extrinsic curvature $K$\,. 

\medskip

The real constant parameter $\eta$ measures the deformation. 
We assume that $\eta$ is positive. In the $\eta \to 0$ limit, 
the classical action (\ref{sinh action}) reduces to the original JT model 
\cite{Jackiw,Teitelboim} (without matter fields)
\begin{eqnarray}
S^{(\eta =0)}_{\Phi } &=& \frac{1}{16\pi G}\int\!\dd^2 x\, \sqrt{-g}\, 
\Phi^2 \left[ R +  \frac{2}{L^2} \right] 
+\frac{1}{8 \pi G}\int\!\dd t\, \sqrt{-\gamma_{tt}}\,\Phi^2 K\,.
\end{eqnarray}
Thus the model (\ref{sinh action}) can be regarded as a deformation of the JT model. 

\medskip

In the following, we will work with the metric in conformal gauge, 
\begin{eqnarray}
\dd s^2 &=&g_{\mu\nu}\dd x^\mu \dd x^\nu= -{\rm e}^{2\omega}\dd x^+ \dd x^-\,,
\end{eqnarray}
where the light-coone coordinates $x^\pm$ are defined as
\begin{eqnarray}
x^\pm &\equiv& t\pm z\,.
\end{eqnarray}
By taking variations of $S_{\Phi}$, the equations of motion are obtained as follows: 
\begin{eqnarray}
4\partial_+\partial_-\Phi^2 +\frac{  {\rm e}^{2\omega}  }{\eta L^2} \sinh\left( 2 \eta\, \Phi^2\right)
&=&0\,,  \label{EOMsinh}\\
4\partial_+\partial_-\omega +\frac{ {\rm  e}^{2\omega}}{ L^2 } 
\cosh\left( 2 \eta\, \Phi^2\right) &=& 0 \,, \label{EOMcosh} \\
- {\rm e}^{2\omega}\partial_+({\rm e}^{-2\omega}\partial_+\Phi^2) 
&=& 0, \label{original const+}\\ 
- {\rm e}^{2\omega}\partial_-({\rm e}^{-2\omega}\partial_-\Phi^2) 
&=&0\,. 
\label{original const-}
\end{eqnarray}

\medskip 

To solve the equations of motion [(\ref{EOMsinh})-(\ref{original const-})] systematically, 
it is very useful to introduce a set of new variables $\omegati_1$ and $\omegati_2$:
\begin{eqnarray}
\omegati_1 \equiv \omega+\eta\,\Phi^2,\qquad \omegati_2 \equiv \omega-\eta\,\Phi^2\,. \label{omegati}
\end{eqnarray}
From (\ref{EOMsinh}) and (\ref{EOMcosh}), one can obtain two Liouville equations for 
$\omegati_1$ and $\omegati_2$~: 
\begin{eqnarray}
4\partial_+\partial_- \omegati_1 + \frac{1}{L^2} {\rm e}^{2 \omegati_1 } &=&0\,, \no\\
4\partial_+\partial_- \omegati_2 +\frac{1}{L^2} {\rm e}^{2\omegati_2 } &=&0\,.  
\label{EOMLiouville}
\end{eqnarray}
It is well known that the general solutions to the Liouville equation are given 
by arbitrary holomorphic functions $X_i^+(x^+)~(i=1,2)$ and anti-holomorphic functions $X_j^-(x^-)~(j=1,2)$:
\begin{eqnarray}
{\rm e}^{2 \omegati_1 }=\frac{4L^2 \partial_+ X_1^+ \partial_- X_1^- }{\left( X_1^+ - X_1^- \right)^2}\,, 
\no\\
{\rm e}^{2 \omegati_2 }=\frac{4L^2 \partial_+ X_2^+ \partial_- X_2^- }{\left( X_2^+ - X_2^- \right)^2}\,. 
\label{Liouville sol}
\end{eqnarray}
By using $X_1^\pm$ and $X_2^\pm$\,, 
the constraint conditions (\ref{original const+}) and (\ref{original const-}) can be simplified into 
\begin{eqnarray}
&&{\rm Sch}\{ X_1^\pm, x^\pm\} - {\rm Sch}\{ X_2^\pm, x^\pm\} =0\,. 
\end{eqnarray}
Here ${\rm Sch}\{X,x\}$ is the Schwarzian derivative defined as 
\begin{eqnarray}
&&{\rm Sch}\{ X, x \} \equiv \frac{ X\/'''}{X\/'} - \frac{3}{2} \left( \frac{ X\/'' }{ X\/' } \right)^2\,. 
\end{eqnarray}
As a result, the metric ${\rm e}^{2 \omega}$ and the dilaton $\Phi^2$ are represented 
by two solutions to the Liouville equations: 
\begin{eqnarray}\label{relation12}
{\rm e}^{ 2 \omega } &&=\sqrt{{\rm e}^{ 2\omegati_{1}} {\rm e}^{2 \omegati_{2 } }} 
= 4L^2\sqrt{ \frac{ \partial_+ X_1^+ \partial_- X_1^- }{  \left( X_1^+ - X_1^- \right)^2}\,
\frac{ \partial_+ X_2^+ \partial_- X_2^- }{  \left( X_2^+ - X_2^- \right)^2}\, }\,,  \\
\Phi^2&&=\frac{\omegati_1-\omegati_2}{2\eta}
=\frac{1}{4 \eta} \log \Biggl|\frac{ \partial_+ X_1^+ \partial_- X_1^- }{\left( X_1^+ - X_1^- \right)^2}\, 
\frac{ \left( X_2^+ - X_2^- \right)^2 }{ \partial_+ X_2^+ \partial_- X_2^- } \Biggr|\,. 
\label{dilaton}
\end{eqnarray}
Notably, the general solutions (\ref{sinh action}) are discussed as 
Yang-Baxter deformations of AdS$_2$\cite{KOYmatter}.

\subsection{Deformed black hole with conformal matter}

Let us next consider to add a conformal matter $\chi$\,, which couples to the Ricci scalar and the dilaton. 
The classical action is given by 
\begin{eqnarray}
S_{\chi} &=&  -\frac{N}{ 24 \pi }\int\!\dd^2 x\, 
\sqrt{-g}\,\Bigl[\chi ( R -2 \eta \nabla^2  \Phi^2) 
+ (\nabla \chi)^2  \Bigr]
-\frac{N}{12 \pi }\int \dd t\, \sqrt{-\gamma_{tt} }\,\chi\,K\,\label{deformedAPchi}\,.
\end{eqnarray}
Here $N$ denotes the central charge of $\chi$. 

\medskip 

Similarly to the derivation of (\ref{EOMLiouville})\,, 
the equations of motion can be rewritten in terms of $\omegati_1$ and $\omegati_2$\,. 
After all, the equations of motion are given by 
\begin{eqnarray}
 \partial_+ \partial_-(\tilde{\omega}_1 + \chi) &=& 0\,, \no\\
4\partial_+\partial_- \tilde{\omega}_{1} + {\rm e}^{2\tilde{\omega}_{1} } 
&=&\frac{16}{3}GN  \eta\, \partial_+ \partial_- \chi \,, \no\\
4\partial_+\partial_- \tilde{\omega}_{2} + {\rm e}^{2\tilde{\omega}_{2} } &=&0 \,,  \no\\
{\rm e}^{\tilde{\omega}_{1} } \partial_{\pm}\partial_{\pm} {\rm e}^{ -\tilde{\omega}_{1} } 
- {\rm e}^{\tilde{\omega}_{2} } \partial_{\pm}\partial_{\pm} {\rm e}^{ -\tilde{\omega}_{2} } 
&=&\frac{2}{3}GN (-\partial_{\pm} \partial_{\pm} \chi + \partial_{\pm} \chi  \partial_{\pm} \chi  
+2  \partial_{\pm} \chi  \partial_{\pm} \tilde{\omega}_{1} )\,. \label{eom-set}
\end{eqnarray}
Note here that the third equation is still the Liouville equation, 
while the second one acquires the source term due to the presence of the conformal matter. 
The last one gives rise to the constraint conditions for the solutions.

\subsubsection*{A black hole solution}

Here, let us derive a black hole solution. Suppose that the solution is static. 
Then, by solving the first equation 
in the set of equations of motion (\ref{eom-set}), $\chi$ can be expressed as\footnote{
The general solution is given by $\chi + \tilde{\omega}_1 = c_1 (x^+-x^-) + c_0$\,, 
where $c_1$ and $c_0$ are arbitrary real constants. We have set that $c_0 =0$ 
for simplicity and $c_1 = \sqrt{\mu}$ 
for later convenience, where $\mu$ is a real positive constant. } 
\begin{eqnarray}
\chi = -\tilde{\omega}_1-\sqrt{\mu}\,(x^+-x^-)\,.
\end{eqnarray}
By eliminating $\chi$ from the other equations, one can derive a couple of Liouville equations 
and the constraint conditions:
\begin{eqnarray}
4\left(1+\frac{4}{3}GN \eta\right)\partial_+\partial_- \tilde{\omega}_{1} 
+ {\rm e}^{2\tilde{\omega}_{1} } &=&0\,, \no\\
4\partial_+\partial_- \tilde{\omega}_{2} + {\rm e}^{2\tilde{\omega}_{2} } &=&0 \,, \no\\
\left(1+\frac{2}{3}GN\right) {\rm e}^{\tilde{\omega}_{1} } \partial_{\pm}\partial_{\pm} 
{\rm e}^{ -\tilde{\omega}_{1} } - {\rm e}^{\tilde{\omega}_{2} } \partial_{\pm}\partial_{\pm} 
{\rm e}^{ -\tilde{\omega}_{2} } &=&\frac{2}{3}GN\mu \,. 
\end{eqnarray}

Now $\tilde{\omega}_{1}$ and $\tilde{\omega}_{2}$ are the general solutions to the Liouville equation: 
\begin{eqnarray}
{\rm e}^{2 \tilde{\omega}_{1} }&=&  \frac{ 4 L^2 ( 1+\frac{4}{3}GN \eta ) }{  \left( X^+_1 - X^-_1 \right)^2} 
\,\partial_+ X^+_1 \partial_- X^-_1\,,  \no\\
{\rm e}^{2 \tilde{\omega}_{2} }&=&   \frac{ 4 L^2 }{  \left( X^+_2 - X^-_2 \right)^2}
\,\partial_+ X^+_2 \partial_- X^-_2\,. 
\label{omegati with c-matter}
\end{eqnarray}
By using $X^\pm_i$ and the Schwarzian derivative, the constraint conditions can be rewritten as
\begin{eqnarray}
\left(1+\frac{2}{3}GN\right){\rm Sch}\{ X^+_1, x^+\} - {\rm Sch}\{ X^+_2, x^+\} 
&=&-\frac{4}{3}GN\mu\,, \no\\
\left(1+\frac{2}{3}GN\right){\rm Sch}\{ X^-_1, x^-\} - {\rm Sch}\{ X^-_2, x^-\} 
&=&-\frac{4}{3}GN\mu\,. 
\end{eqnarray}
It is an easy task to see that 
\[
X^\pm_{1,2}=\tanh({\sqrt{\mu}\,x^\pm})
\] 
satisfy the constraint conditions.
In general, the linear fractional transformations of them,
\begin{eqnarray}
X^\pm_{1,2} &=&\frac{a \tanh(\sqrt{\mu}\, x^\pm) +b}{c \tanh(\sqrt{\mu}\, x^\pm) +d}\,, \label{coor}
\end{eqnarray}
also satisfy the constraint conditions, thanks to a property of the Schwarzian derivative. 

\medskip

By taking certain parameters, a deformed black hole solution \cite{KOY,KOYmatter} is given by 
\begin{eqnarray}
{\rm e}^{2 \omega }&=&\frac{ 4\mu L^2 (1-\eta^2\mu)\sqrt{1+\frac{ 4}{3}GN \eta} }{  
\sinh^2(2 \sqrt{\mu}\,Z )-\eta^2\mu \cosh^2(2\sqrt{\mu}\,Z)}\,, 
\label{dBHmetric} \\
\Phi^2 &=& \frac{1}{2 \eta} \log \left|
\frac{1+\eta\sqrt{\mu}\, \coth(2\sqrt{\mu}\,Z)}{1-\eta\sqrt{\mu}\, 
\coth(2\sqrt{\mu}\, Z)}\right|+\frac{1}{4\eta}\log\left(1+\frac{ 4}{3}GN \eta \right)\,,  
\label{dBHdilaton}
\end{eqnarray}
where the range of $\sqrt{\mu}$ is restricted as 
\[
0 \leq \sqrt{\mu} \leq \frac{1}{\eta} 
\]
so as to ensure the positivity of the exponential in (\ref{dBHmetric})\,.
Notably, the solutions have a naked singularity intrinsic to the deformation. 
For the detail, see \cite{KOY,KOYmatter}.

\section{Moving to a proper frame: Weyl transformation}

In this section, we shall introduce a new proper frame, which was originally utilized 
by Frolov and Zelnikov \cite{Frolov}. 
One can see that in this proper frame, the deformed JT model can be recaptured 
as a Liouville dilaton gravity model 
with a cosmological constant term, while solutions are still given by $\omegati_{1}$ 
and $\omegati_{2}$\,.

\medskip 

The proper frame is introduced through a dilaton-dependent Weyl transformation: 
\begin{eqnarray}
g_{\mu\nu}={\rm e}^{-2 \eta\, \Phi^2}\tilde{g}_{\mu\nu}\,.  
\label{new frame}
\end{eqnarray}
In conformal gauge, $\omegati_1$ plays the role of the conformal factor in front of the metric: 
\begin{eqnarray}
\dd \tilde{s}^2 = \tilde{g}_{\mu\nu}\dd x^\mu \dd x^\nu = -{\rm e}^{2 \tilde{\omega}_1}\dd x^+ \dd x^-\,.
\end{eqnarray}
In terms of the new metric $\tilde{g}_{\mu\nu}$, the classical action of the deformed JT model 
(\ref{sinh action}) can be rewritten into the following form: 
\begin{eqnarray}
\tilde{S}_{\Phi } &=& \frac{1}{16\pi G}\int\dd^2 x \sqrt{-\tilde{g} }\left[
\Phi^2 \tilde{R} -2\eta (\tilde{\nabla} \Phi^2)^2 
- \frac{1}{2\eta L^2 }\left( {\rm e}^{-4\eta \Phi^2} -1\right)\right]\,.\label{Liouville action}
\end{eqnarray}
Note here that the kinematic term of $\Phi^2$ is well-defined because we have assumed that 
$\eta $ is a positive real constant. The potential is now bounded from below, 
but it is a run-away type potential.  It is also remarkable that $\Phi^2$ (instead of $\Phi$)  
appears in the classical action (\ref{Liouville action}) and $\Phi^2$ should be definitely positive. 
Hence this is not the usual Liouville gravity but rather a {\it constrained} Liouville gravity. 
Interestingly, this constrained system can also be derived 
from Einstein-Hilbert action with a cosmological constant \cite{GJ1}.

\medskip 

It is remarkable that the equations of motion for $\omega$ and $\Phi^2$ are equivalent 
to the equations for $\omegati_1$ and $\omegati_2$\,. 
Thus, the solutions to (\ref{Liouville action}) are obtained 
by $\omegati_1$ and $\omegati_2$ as in  (\ref{Liouville sol})\,.
From $\omegati_1$ and $\omegati_2$, the dilation is determined by (\ref{dilaton}) again.  
However, in the proper frame, the metric is given by $\omegati_1$ only.

\medskip

In summary, the general vacuum solutions to (\ref{Liouville action}) are given by
\begin{eqnarray}
{\rm e}^{2 \omegati_1 }&=&\frac{4L^2 \partial_+ X_1^+ \partial_- X_1^- }{ 
\left( X_1^+ - X_1^- \right)^2}\,, 
\no\\
\Phi^2&=&\frac{1}{4 \eta} \log \Biggl|\frac{ \partial_+ X_1^+ \partial_- X_1^- }{ 
\left( X_1^+ - X_1^- \right)^2}\, 
\frac{ \left( X_2^+ - X_2^- \right)^2 }{ \partial_+ X_2^+ \partial_- X_2^- } \Biggr| \,. 
\end{eqnarray}
Note here that the metric is given by the general solution to Liouville equation 
and the rigid AdS$_2$ geometry is preserved in the new frame (i.e., proper frame). 
This result indicates that the Weyl transformation carried out here has undone 
the Yang-Baxter deformation of the metric, while the deformation effect is now encoded 
into only the dilaton part. It should be remarked that this is a rather natural result, 
noticing that the Yang-Baxter deformation effect can be factored out as the overall factor of the metric, 
as shown in \cite{KOY,KOYmatter}.

\section{A new black hole solution and its thermodynamics}

In this section, we present a new black hole solution with a conformal matter\footnote{
This solution is different from the one introduced in Sec.\ 2.2. For the detail, see Appendix A.}. 
The proper frame (\ref{new frame}) enables us to construct an AdS$_2$ black hole solution 
(i.e., the metric is the same as the undeformed case \cite{AP})\footnote{By performing 
the inverse Weyl transformation for the new solution, a naked singularity appears again. 
In some sense, this Weyl transformation is similar to a coordinate transformation,
 which was proposed in \cite{Kame}, in the case of the $\eta$-deformed AdS$_5$ \cite{ABF}.}.  
Then, we compute the entropy of the black hole solution in two manners: 
1) the Bekenstein-Hawking entropy and 2) the boundary stress tensor with a certain counter-term. 
Both results are consistent. 

\subsubsection*{A new black hole solution}

In the proper frame, the classical action of the matter ($\ref{deformedAPchi}$) is given by
\begin{eqnarray}
\tilde{S}_{\chi} &=&  -\frac{N}{ 24 \pi }\int\!\dd^2 x\, \sqrt{-\tilde{g}}\,\left[
\chi  \tilde{R}  + (\tilde{\nabla} \chi)^2  \right] 
-\frac{N}{12 \pi }\int\! \dd t\, \sqrt{-\tilde{\gamma}_{tt}}\,\chi\,K\,. 
\label{proper-matter}
\end{eqnarray}
Note here that in the proper frame, $\chi$ couples to only the Ricci scalar, 
while in the old frame (\ref{deformedAPchi}), $\chi$ coupled to both Ricci scalar and dilaton. 
This point is the same as in the undeformed case \cite{AP}. 

%

\medskip

In order to find out a black hole solution of the system (\ref{proper-matter}),  
let us take black hole coordinates 
for $X^\pm_{1}$ as  
\begin{eqnarray}
X^\pm_1(x^\pm) &=&\frac{1}{\sqrt{\mu}}\tanh(\sqrt{\mu}x^\pm ) \,, 
\end{eqnarray}
by following \cite{AP}. 
Then, for $X^\pm_2$, we take the following linear fractional transformation: 
\begin{eqnarray}
X^+_2(x^+) &=& \frac{  \tanh(\sqrt{\mu}x^+)+ 2 \eta\sqrt{\mu} }{2\eta \mu \tanh(\sqrt{\mu}x^+) +\sqrt{\mu} }\,, \no\\
X^-_2(x^-) &=& \frac{1}{\sqrt{\mu}}\tanh(\sqrt{\mu}\,x^- )\,. 
\end{eqnarray}
Thus we obtain the static solutions for $\omegati_1$ and $\omegati_2$~: 
\begin{eqnarray}
{\rm e}^{2 \tilde{\omega}_1 } &=& \frac{ 4\mu (1+\frac{4}{3}GN \eta)L^2 }{
\sinh^2(2 \sqrt{\mu}Z )}\,, \\
{\rm e}^{2 \tilde{\omega}_2 } &=& \frac{ 4\mu ( 1-4\eta^2\mu )L^2 }{ 
\left( \sinh(2 \sqrt{\mu}Z ) +2\eta \sqrt{\mu} \cosh(2\sqrt{\mu} Z) \right)^2 }\,. 
\label{4.5}  
\end{eqnarray}
Hence a black hole solution with a conformal matter is 
\begin{eqnarray}
{\rm e}^{2 \tilde{\omega}_1 }&=&\frac{ 4\mu (1+\frac{4}{3}GN \eta)L^2 }{\sinh^2(2 \sqrt{\mu}Z )}\,, 
\label{newBHmetric}\\
\Phi^2 &=& \frac{1}{2 \eta} \log \Big|
1+2 \eta \sqrt{\mu}\, \coth(\sqrt{\mu}\,Z)\Big|+\Phi_0^2\,.  
\label{newBHdilaton}
\end{eqnarray}
Here $\Phi_0^2$ is the constant part of the dilaton given by 
\begin{eqnarray}
\Phi_0^2 &\equiv& \frac{1}{4\eta} \log \left( \frac{1+\frac{4}{3}GN \eta}{ 1-4\eta^2\mu } \right)\,.  
\label{Phi0}
\end{eqnarray}
You see that the matter contribution just rescales the metric and shifts the dilaton by a constant. 
Note here that the allowed region of $\mu$ is restricted like
\begin{equation}
0 \leq \sqrt{\mu} \leq \frac{1}{2\eta}\, \label{range}
\end{equation}
so as to make the value of $\Phi_0^2$ well-defined and preserve the positivity of (\ref{4.5})\,.  

\medskip 

It should be remarked that this solution is different from the previous black hole solution 
with (\ref{dBHmetric}) and (\ref{dBHdilaton}), though the two solutions are quite similar 
but the $\mu$-dependence of the metric and the range of $\sqrt{\mu}$ are different 
as explicitly shown in Appendix A. 

\medskip 

By taking the undeformed limit $\eta \to 0$\,, this solution reduces to the black hole solution 
with a conformal matter in the undeformed case \cite{AP}:  
\begin{eqnarray}
{\rm e}^{2 \omega }&=&\frac{ 4\mu L^2 }{  \sinh^2(2 \sqrt{\mu}Z )}\,, \\
\Phi^2 &=& \sqrt{\mu}\, \coth(\sqrt{\mu}\,Z)+\frac{1}{3}GN\,. 
\end{eqnarray}

\medskip

In the following, let us evaluate the black hole entropy associated with (\ref{newBHmetric}) 
and (\ref{newBHdilaton}) in two manners.

\subsubsection*{1) Bekenstein-Hawking entropy }

Let us first compute the Bekenstein-Hawking entropy. 
From the black hole metric (\ref{newBHmetric}), the Hawking temperature is computed as 
\begin{eqnarray}
T_{\rm H}=\frac{ \sqrt{\mu} }{\pi}\,.  
\label{HT}
\end{eqnarray}
This is the same as in the undeformed case \cite{AP}. 
From the classical action, one can read off the effective Newton constant $G_{\rm eff}$ as 
\begin{eqnarray}
\frac{1}{G_{\rm eff}} = \frac{\Phi^2}{G} -\frac{2 N \chi}{3} \,. 
\label{Geff}
\end{eqnarray}
Given that the horizon area A is $1$,  
the Bekenstein-Hawking entropy $S_{\rm BH}$ is evaluated as 
\begin{eqnarray}
S_{\rm BH} 
&=& \left. \frac{A}{4 G_{\rm eff} }\right|_{Z \to \infty} \,\no\\
&=& \frac{ \text{arctanh}{ (2 \pi \,T_{\rm H} \, \eta)}  }{8 G \eta} +\frac{N}{6}\log(T_{\rm H}) 
+ {\rm constant}\,. 
\label{BH}
\end{eqnarray}
The last term is a constant term independent of the Hawking temperature. 
{Note here that the argument of arctanh should be less than 1. This means that 
\[
0 \leq  T_{\rm H} \leq \frac{1}{2\pi \eta}\,.
\]
This range agrees with the possible values of $\sqrt{\mu}$ given in (\ref{range})\,.

\subsubsection*{2) Boundary stress tensor}

In conformal gauge, the total action, including the Gibbons-Hawking term, is given by 
\begin{eqnarray}
\tilde{S}_{\Phi} 
&=&\frac{1}{8 \pi G}\int\!\dd^2 x \left[
-4 \partial_{(+} \Phi^2 \partial_{-)} \tilde{\omega}_1 +4\eta\partial_+\Phi^2\partial_-\Phi^2 
-  \frac{1}{4\eta L^2} {\rm e}^{2 \tilde{\omega}_1 }({\rm e}^{-4\eta\Phi^2}-1) \right]\,, 
\no\\
\tilde{S}_{\chi}
&=&\frac{N}{6 \pi}\int\!\dd^2 x \,\Bigl[
\partial_+\chi\partial_-\chi+2\partial_{(+}\chi\partial_{-)}\tilde{\omega}_1\Bigr]\,.  
\end{eqnarray}
By using the explicit expression of the black hole solution, 
the on-shell bulk action can be evaluated on the boundary.

\medskip

The on-shell action diverges at the boundary $Z=0$\,, hence one needs to 
introduce a cut-off as $Z=\epsilon~(>0)$\,, where $\epsilon$ is an infinitesimal quantity.  
Then the on-shell action can be expanded with respect to $\epsilon$ like   
\begin{eqnarray}
\tilde{S}_{\Phi}+\tilde{S}_{\chi} &=& \int \! \dd t\, \left[
\frac{1+\frac{4}{3}GN\eta }{ 16 \pi G\, \eta\, \epsilon  }+O(\epsilon^1) \right]\,. 
\end{eqnarray}
Here we have ignored the terms which vanish in the $\epsilon \to 0$ limit, 
and only the divergent term has explicitly been written down.  
To cancel out the divergence, it is necessary to add an appropriate counter-term. 

\medskip 

Our proposal for the counter-term is the following:\footnote{The uniqueness of 
the counter-term has not been confirmed.  
It is significant to revisit it by following the works \cite{Cvetic,GJ2,GJ3}. } 
\begin{eqnarray}
\tilde{S}_{\rm ct} &=& \int\!\dd t\, \frac{\sqrt{-\tilde{\gamma}_{tt}} }{L'}\, 
\left[ \frac{-1}{16\pi G\,\eta } 
\Big(1+(2\eta\,\Phi_0^2-1){\rm e}^{-2\eta(\Phi^2-\Phi_0^2)} \Big) -\frac{N}{24\pi} \right]\,. 
\label{counter2}
\end{eqnarray}
Here $\Phi_0$ is the constant defined in (\ref{Phi0}) and $L'$ is the rescaled AdS radius defined as 
\begin{eqnarray}
L'^2 \equiv L^2 \left(1+\frac{ 4}{3}GN \eta \right)\,. 
\end{eqnarray}
Then the extrinsic metric $\tilde{\gamma}_{tt}$ on the boundary is defined as 
\[
\tilde{\gamma}_{tt} \equiv \left. - {\rm e}^{2\tilde{\omega}_1 } \right|_{Z=\epsilon} \,. 
\]
Note here that the counter-term (\ref{counter2}) is local and is represented basically by the Liouville potential.  
In the undeformed limit $\eta \to 0$\,, the counter-term (\ref{counter2}) reduces to
\begin{eqnarray}
\tilde{S}_{\rm ct}^{(\eta =0)} &=& \int\!\dd t\, \frac{\sqrt{-\gamma_{tt}}}{L} \, 
\left( -\frac{\Phi^2}{8 \pi G} -\frac{N}{24 \pi }\right)\,. 
\end{eqnarray}
This is nothing but the counter-term utilized in the undeformed case \cite{AP}.

\medskip

It is straightforward to check that the sum $\tilde{S} = \tilde{S}_{\Phi} + \tilde{S}_{\chi} 
+ \tilde{S}_{\rm ct}$ 
becomes finite on the boundary by using the expanded form of the counter-term (\ref{counter2}): 
\begin{eqnarray}
\tilde{S}_{\rm ct} &=& \int \! \dd t\, \left[
 -\frac{1+\frac{4}{3}GN\eta }{16 \pi G \eta\, \epsilon } + \frac{1-2\eta\Phi_0^2 }{ 16 \pi G\, \eta^2}\, 
+O(\epsilon) \right]\,.\no
\end{eqnarray}
In a region near the boundary, the warped factor of the metric can be expanded as 
\begin{eqnarray}
 {\rm e}^{2 \omegati_1} &=& \frac{L'^2}{ \epsilon^2 } + O(\epsilon^0)\,. 
\end{eqnarray}
Hence, by normalizing the boundary metric as 
\[
\hat{\gamma}_{tt}  =\frac{ \epsilon^2 }{L'^2} \, \tilde{\gamma}_{tt}\,, 
\]
the boundary stress tensor is defined as 
\begin{eqnarray}
\langle \hat{T}_{tt} \rangle 
&\equiv& \frac{-2}{\sqrt{-\hat{\gamma}_{tt}}}\, \frac{\delta S}{\delta \hat{\gamma}^{tt} } 
= \lim_{ \epsilon \to 0 } \frac{\epsilon}{L'}\, \frac{-2}{\sqrt{-\tilde{\gamma}_{tt} } }\, 
\frac{\delta S}{\delta \tilde{\gamma}^{tt} }\,.  
\end{eqnarray}
After all, $\langle \hat{T}_{tt} \rangle$ is evaluated as  
\begin{eqnarray}
\langle \hat{T}_{tt} \rangle &=& -\frac{\log(1-4\eta^2 \mu) }{32 \pi G\, \eta^2} 
+ \frac{ \log \left( 1+\frac{4}{3}GN\eta \right) }{ 32 \pi G\, \eta^2 }+\frac{N \sqrt{\mu}}{6\pi}\,. 
\end{eqnarray}
This expression of $\langle \hat{T}_{tt} \rangle$ should be identified with thermodynamic energy $E$ like 
\begin{eqnarray}
E &=& -\frac{ \log(1-4 \pi^2 T_{\rm H}^2 \eta^2 ) }{32 \pi G\, \eta^2} 
+ \frac{ \log \left( 1+\frac{4}{3}GN\eta \right) }{ 32 \pi G\, \eta^2 }+\frac{N}{6}T_{\rm H}\,,
\end{eqnarray}
where we have used the expression of the Hawking temperature (\ref{HT})\,.  

\medskip 

Then, by solving the thermodynamic relation,
\begin{eqnarray}
\dd E &=&\frac{\dd S}{T_{\rm H} }\,,
\end{eqnarray}
the entropy is obtained as  
\begin{eqnarray}
S &=& \frac{ \text{arctanh}{(2 \pi T_{\rm H} \eta)} }{ 8 G \eta } 
+ \frac{N}{6}\log(T_{\rm H})+S_{T_{\rm H}=0}\,.
\end{eqnarray}
Here $S_{T_{\rm H}=0}$ has appeared as an integration constant that measures 
the entropy at zero temperature. 
Thus the resulting entropy is consistent with the Bekenstein-Hawking entropy 
(\ref{BH})\,, up to the temperature-independent constant.

\section{Conclusion and discussion}

In this paper we have revisited a deformed Jackiw-Teitelboim model 
with a hyperbolic dilaton potential, which was constructed 
in the preceding work \cite{KOY,KOYmatter}. 
By employing a Weyl transformation to this deformed model, we have discussed 
a Liouville type potential with a cosmological constant term. 
Then we have found regular solutions coupled to a conformal matter 
by using $SL(2)$ transformations. 

\medskip 

For a black hole solution, its thermodynamic behavior has been investigated.  
In particular, the Bekenstein-Hawking entropy can be reproduced by evaluating the boundary stress 
tensor with a local counter-term, which is essentially provided 
by a Liouville type potential as in the undeformed case \cite{AP}. 
Notably, this counter-term is concise, 
in comparison to the previous ones utilized in \cite{KOY,KOYmatter}. 

\medskip 

It would be significant to argue the implication of the Weyl transformation utilized 
in our analysis in the context of Yang-Baxter deformation. 
An important observation is that the deformation is nothing but the overall factor 
in front of the AdS$_2$ metric. Hence the undeformed AdS$_2$ may be realized by performing an 
appropriate Weyl transformation.  After all, the vestiges of the deformation is encoded into the dilaton 
and other matter fields. Of course, in higher dimensions, 
it would not be possible to realize the undeformed metric completely. 
But it should be interesting to concentrate on the AdS$_2$ subregion of 
higher-dimensional Yang-Baxter deformed backgrounds. 
In particular, Yang-Baxter deformed backgrounds suffer from a naked singularity. 
A famous example is the $\eta$-deformed background \cite{ABF} with a classical $r$-matrix 
of the Drinfeld-Jimbo type and 
an $\eta$-deformed AdS$_2\times$S$^2$ is discussed in \cite{HRT,LRT}. 
As an exercise, it would be nice to apply our argument to 
this $\eta$-deformed AdS$_2\times$S$^2$ case. 
We will report some results on this issue in the near future \cite{future}.

\subsection*{Acknowledgments}

We are very grateful to Masafumi Fukuma, Hikaru Kawai, Hideki Kyono and Jun-ichi Sakamoto 
for valuable comments and useful discussions.  We appreciate Daniel Grumiller for letting us 
know significant references.
The work of K.Y. was supported by the Supporting Program for Interaction-based Initiative 
Team Studies (SPIRITS) from Kyoto University, a JSPS Grant-in-Aid for Scientific Research (B) 
No.\,18H01214 and a JSPS Grant-in-Aid for Scientific Research (C) No.\,15K05051.
This work is also supported in part by the JSPS Japan-Russia Research Cooperative Program.

\appendix

\section*{Appendix}

\section{Comparison of two black hole solutions }

In Sec.\ 4, we have considered a black hole solution in the proper frame. 
This solution is slightly different from the one presented in Sec.\ 2.2 
(or equivalently found in \cite{KOY,KOYmatter}). 
In the following, we show the difference explicitly 
by comparing the two solutions after mapping the previous solution in Sec.\ 2.2 
to the proper frame. 

\subsection{The deformed black hole solution in Sec.\ 2.2}

First of all, let us revisit the black hole solution [(\ref{dBHmetric}) and (\ref{dBHdilaton})]. 
In the following, we will focus upon the case without matters for simplicity (i.e., the case with 
$N=0$).  

\medskip 

In Sec.\ 2.2, we employed the following linear fractional transformation: 
\begin{eqnarray}
X^+_1(x^+) &=& \frac{ (1-\eta \beta)X^+(x^+) - 2 \eta \alpha }{-2 \eta \gamma X^+(x^+) 
+ (1+\eta \beta) } \,, 
\qquad X^-_1(x^-)  = X^-(x^-)\,, \no \\
X^+_2(x^+) &=& \frac{ (1+\eta \beta)X^+(x^+) + 2\eta \alpha }{2\eta \gamma X^+(x^+) 
+ (1-\eta \beta) }\,,  
\qquad X^-_2(x^-) = X^-(x^-)\,. 
\label{YB form}
\end{eqnarray}
Then, $\omegati_1$ and $\omegati_2$ are given by
\begin{eqnarray}
{\rm e}^{2 \omegati_{1} }=\frac{4 \left( 1-\eta^2(\beta^2 +4 \alpha \gamma) \right)
\partial_+ X^+\partial_- X^- }{ 
\left( X^+ -X^- - \eta( 2 \alpha +\beta (X^+ +X^-) -2\gamma X^+ X^-) \right)^2 }\,, \no\\
{\rm e}^{ 2 \omegati_{2} }=\frac{4\left( 1-\eta^2(\beta^2 +4 \alpha \gamma) \right) 
\partial_+ X^+\partial_- X^-}{ 
\left( X^+ -X^- + \eta(2 \alpha +\beta (X^+ +X^-) -2\gamma X^+ X^-) \right)^2 }\,. 
\end{eqnarray}
In particular, by taking the following parameters 
\begin{equation}
\alpha=\frac{1}{2}\,, \quad \beta=0\,, \quad \gamma=\frac{\mu}{2}
\label{BHparameter}
\end{equation}
and the black hole coordinates 
\begin{equation}
X^\pm=\frac{1}{\sqrt{\mu}}\tanh\left[\sqrt{\mu}\,(T \pm Z)\right]\,, \label{BH-coords}
\end{equation}
the deformed blach hole solution [(\ref{dBHmetric}) and (\ref{dBHdilaton})]  
can be reproduced.

\subsection{Mapping to the proper frame}

The next task is to map the solution [(\ref{dBHmetric}) and (\ref{dBHdilaton})]  
to the proper frame. 

\medskip 

In order to realize the rigid AdS$_2$ metric in the proper frame, 
we need to perform the following coordinate transformation:
\begin{eqnarray}
X^+(x^+) = \frac{ (1-\eta\,\beta) \tilde{X}^++2\eta\,\alpha}{2\eta\,\gamma
\tilde{X}^+ +(1-\eta\,\beta)}\,, 
\qquad X^-(x^+)&=&\tilde{X}^-\,.
\end{eqnarray}
Then the holomorphic functions are transformed as 
\begin{eqnarray}
X^+_1 = \tilde{X}^+\,,\qquad 
X^+_2 = \frac{\bigr((1+\eta\,\beta)^2+4\eta^2\,\alpha\,\gamma\bigl) 
\tilde{X}^++4\eta\,\alpha}{4\eta\,\gamma\tilde{X}^+ +\bigr((1-\eta\,\beta)^2+4\eta^2
\,\alpha\,\gamma\bigl)}\,.
\end{eqnarray} 
Again, by choosing the parameters (\ref{BHparameter}) 
and the black hole coordinates (\ref{BH-coords})\,, 
the black hole solution in the proper frame can be obtained as 
\begin{eqnarray}
{\rm e}^{2 \omegati_{1} }&&=\frac{4\mu L^2}{\sinh^2(2\sqrt{\mu}\,Z)}\,, \no\\
{\rm e}^{ 2 \omegati_{2} }&&= \left(\frac{1-\eta^2\mu}{1+\eta^2\mu}\right)^2
\frac{4\mu L^2}{ \left( \sinh(2\sqrt{\mu}\,Z) 
+\frac{2\eta\sqrt{\mu}}{1+\eta^2\mu}\cosh(2\sqrt{\mu}\,Z) \right)^2 }\,. 
\end{eqnarray}
The dilaton is given by
\begin{eqnarray}
\Phi^2&&= 
\frac{1}{2\eta}\log\left| 1 +\frac{2\eta\sqrt{\mu}}{1+\eta^2\mu}\coth(2\sqrt{\mu}\,Z) \right|
+\frac{1}{2\eta}\log\left(\frac{1+\eta^2\mu}{1-\eta^2\mu}\right)\,. 
\end{eqnarray}
These expressions are clearly different from the solution with (\ref{newBHmetric}) 
and (\ref{newBHdilaton}) for the $N=0$ case. 

%

\end{document}